\documentclass[review]{elsarticle}
\usepackage{subfigure}
\usepackage{epsfig,graphics,subfigure,psfrag,amsmath,amssymb}
\usepackage{amsmath,amssymb,amsfonts}
\usepackage{algorithmic}
\usepackage{graphicx}
\usepackage{textcomp}
\usepackage[vlined,boxed,ruled]{algorithm2e}
\usepackage{dsfont}
\newtheorem{definition}{Definition}

\journal{Information Sciences}

\begin{document}

\begin{frontmatter}

\title{Achieving Trust-Based and Privacy-Preserving Customer Selection in Ubiquitous Computing}

\author[a]{Chuan Zhang}
\author[a]{Liehuang Zhu\corref{mycorrespondingauthor}}

\cortext[mycorrespondingauthor]{Corresponding author}
\ead{liehuangz@bit.edu.cn}

\author[a]{Chang Xu\corref{mycorrespondingauthor}}

\ead{xuchang@bit.edu.cn}

\author[a]{Kashif Sharif}

\author[b]{Ximeng Liu}

\author[c]{Xiaojiang Du}

\author[d]{Mohsen Guizani}

\address[a]{Beijing Engineering Research Center of Massive Language Information Processing and Cloud Computing Application, School of Computer Science and Technology, Beijing Institute of Technology, Beijing, China.}

\address[b]{School of Information Systems, Singapore Management University and College of Mathematics and Computer Science, Fuzhou University.}

\address[c]{Department of Computer and Information Sciences, Temple University, Philadelphia, USA.}

\address[d]{Department of Electrical and Computer Engineering, University of Idaho, Moscow, Idaho, USA.}

\begin{abstract}
The recent proliferation of smart devices has given rise to ubiquitous computing, an emerging computing paradigm which allows anytime \& anywhere computing possible. In such a ubiquitous computing environment, customers release different computing or sensing tasks, and people, also known as data processors, participate in these tasks and get paid for providing their idle computing and communication resources. Thus, how to select an appropriate and reliable customer while not disclosing processors' privacy has become an interesting problem. In this article, we present a trust-based and privacy-preserving customer selection scheme in ubiquitous computing, called TPCS, to enable potential processors select the customers with good reputation. The basic concept of TPCS is that each data processor holds a trust value, and the reputation score of the customer is calculated based on processors' trust values and feedbacks via a truth discovery process. To preserve processors' privacy, pseudonyms and Paillier cryptosystem are applied to conceal each processor's real identity. In addition, three authentication protocols are designed to ensure that only the valid data processors (i.e., the processors registering in the system, holding the truthful trust values, and joining the computing tasks) can pass the authentication. A comprehensive security analysis is conducted to prove that our proposed TPCS scheme is secure and can defend against several sophisticated attacks. Moreover, extensive simulations are conducted to demonstrate the correctness and effectiveness of the proposed scheme.
\end{abstract}

\begin{keyword}
\texttt{Ubiquitous computing, trust, privacy-preserving, selection}
\end{keyword}

\end{frontmatter}

\section{Introduction}
Nowadays, the fast development of smart devices (e.g., vehicles \cite{ZhuASAP, zhuzxdxsm}, implantable medical devices \cite{HeiD11,HeiDWH10}, wearable devices \cite{Kim2014Energy}) embedded with increasingly powerful computational and communication resources, has given rise to ubiquitous computing, a revolutionary computing paradigm which integrates surrounding devices to provide numerous novel services at anytime, anywhere, and by any means \cite{MaZCCYHJ06}. In ubiquitous computing, all objects (e.g., smart devices, human bodies, wireless sensors) can be considered as computers or data processors. These objects, besides serving themselves, can also participate in different computing tasks released by the resource-constrained customers (e.g., individual users, companies, organizations), and get paid by providing their idle resources. By doing so, ubiquitous computing is able to greatly make use of surrounding idle resources and drastically change the ways we live and obtain services.

Although many benefits can be gained by ubiquitous computing, some new challenges arise. Since the processors will run automatically under the control of the customers, the reliability of customers is particularly important. Some customers may unintentionally or intentionally abuse processors' resources, give unexpected reward, or even put processors in dangerous situations. For instance, by utilizing the computing capacities of surrounding vehicles, traffic problems such as route navigation or environmental monitoring can be resolved. However, some malicious customers may give rewards which do not match processors' workloads, or collect vehicles' private information such as location and driving habits. Thus, it is essential to identify such customers before joining a ubiquitous computing task. Normally, the performances of customers can be judged by processors' feedbacks \cite{HuLZ17, HuLZS17}. However, the problem here is that the feedbacks given by different processors may vary significantly due to different working loads, incomplete views of observations, or even malicious evaluations. When aggregating these feedbacks, traditional methods such as voting or averaging, which treat all processors equally are not suitable.

\begin{figure}
	\centering
	\includegraphics[ width=12cm]{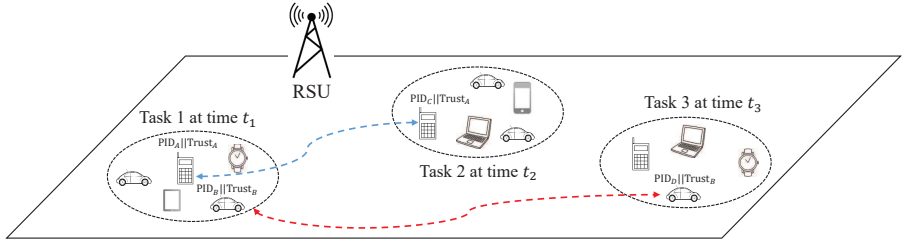}
	\caption{Using trust values to link pseudonyms in a given period of time.}
	\label{fig:linkable}
\end{figure}

An ideal approach to resolve the above challenge is to involve trust values for all processors and make the aggregated reputation scores closed to the feedbacks provided by reliable processors. Nevertheless, another critical issue which must be addressed is the privacy of processors. Although pseudonymous \cite{XuXYH18} and anonymous authentication \cite{HuXHW11, YangXXWLY19} can be used to conceal processor's identity information, processors' location and trajectory privacy may still be disclosed by linking their trust values. To illustrate, we consider a scenario in Fig. \ref{fig:linkable}. At time $t_1$, two processors (i.e., the cell phone and the vehicle) join a task, and at time $t_2$ and $t_3$, they join different tasks respectively. Although their pseudonyms have been changed, their trust values (i.e., $\text{Trust}_A$ and $\text{Trust}_B$) remain unchanged in a certain period of time. By linking their trust values, the trajectories of the processors can be easily reconstructed. Hence, it is important to design a trust-based customers selection scheme which does not sacrifice processors' privacy.

In order to address the above challenges, we present a trust-based and privacy-preserving  customer selection scheme, called TPCS, to rank the customers according to processors' feedbacks while not disclosing their privacy. The general process of TPCS can be described as follows. After finishing a ubiquitous task, processors deliver their feedbacks and encrypted trust values to the roadside units (RSUs), which will collaborate with the Service Provider (SP) to calculate the customer's reputation score using a truth discovery method. During the whole process, processors' privacy will not be disclosed to any other parties. Below, we have summarized the major contributions of this work.
\begin{itemize}
	\item First, we design a filtering truth discovery based evaluation algorithm to process the feedback information received from processors. By this algorithm, the proposed scheme can effectively estimate the performances of customers and processors. This allows optimal selection of customer in a ubiquitous computing environment.
	\item Second, we use pseudonyms and Paillier cryptosystem to protect data processors' privacy. Moreover, three authentication protocols are designed to ensure that only the legitimate processors can pass the authentication. To the best of our knowledge, our work is the first attempt to resolve the security and privacy issues in customers selection in the ubiquitous computing environment.
	\item Third, we conduct a comprehensive security analysis to demonstrate that the scheme presented is not only secure, but can also defend against different sophisticated attacks. Additionally, extensive simulations are performed to validate the correctness \& effectiveness of TPCS.
\end{itemize}

The rest of paper has been organized into 8 sections, where system and threat model along with design goals have been discussed in section 2, followed by preliminary discussion in Section 3. The TPCS scheme functionality is explained in section 5, and its security analysis \& evaluation are presented in section 6. Related works and conclusion are detailed in sections 7 and 8 respectively.

\section{Models and Design Goal}
In order to better present the proposed scheme, we first describe the system model, and give details of the threat model. Based on these, we develop the design goals of our proposed scheme.

\subsection{System Model}
\begin{figure*}[htb]
	\centering
	\includegraphics[width=1\textwidth]{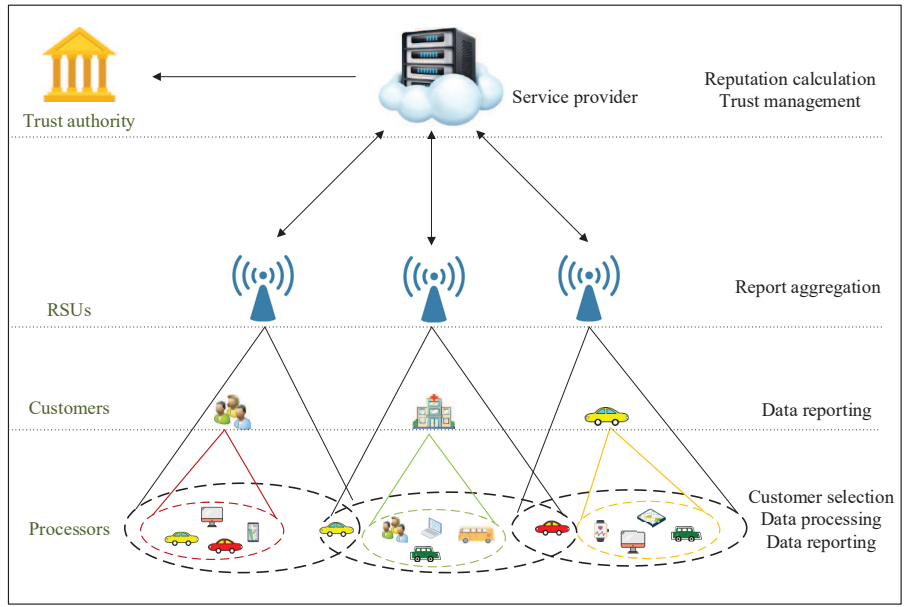}
	\caption{System model.}
	\label{fig:system}
\end{figure*}

The overall model considers a typical scenario of ubiquitous computing. The RSUs are widely deployed in a given area, and all customers and processors can communicate with RSUs through their communication resources. Particularly, the system consists of a trust authority (TA), RSUs, a service provider (SP), customers, and processors, as shown in Fig.~\ref{fig:system}.

\begin{itemize}
	\item Trust Authority (TA): This entity is in charge of all participating parties, and also maintains a database to store processors' trust values. We assume that it is fully capable of storing and performing computation on data generated by other entities. After receiving processors' trust values, it can predict their future behaviors based on historical data.
	\item Service Provider (SP): SP connects all RSUs and stores feedbacks and trust values sent from these roadside units. Upon receiving the data, SP executes a truth discovery based evaluation algorithm to calculate the reputation scores for customers. Similarly, for the query of a potential processor for the task join request, SP can respond it by recommending the customers with high reputation scores.
	\item Roadside Units (RSUs): RSUs are subordinates of SP. They are widely deployed and can cover a wide area. They collect processors' feedbacks and trust values, and then forward them to SP. In particular, they have limited computation capacities, which ensures that they can authenticate processors' identities and perform aggregation operations.
	\item customers: customers can be companies, individual processors, vehicles, and organizations. They have insufficient computation and communication capabilities to perform the tasks by themselves, so they give benefits to hire processors to help them finish their tasks.

	\item Processors: Each processor is embedded with computation and communication units which enable them automatically perform the sensing, computation, and communication tasks received from the customers. After finishing the tasks, each of them will upload its feedback to the nearby RSU for customer evaluation. Besides, the processors update their trust values from TA at regular intervals.

\end{itemize}

\subsection{Threat Model}
TA is fully trusted because it generates the public and private keys for all roles. We assume TA is under strong physical protection and cannot be compromised. SP and RSUs are both considered to be honest but curious. In other words, they will honestly perform the given tasks but try to infer processors' location and trajectory privacy by linking their identities or trust values. Note that, SP and RSU will not collude with each other. This is a common assumption in existing fog-based applications \cite{LuHLG17,XueHMWHY18,ZhangZXSDG19}. The customers are supposed to control the whole ubiquitous computing task. However, their performances may vary differently. For any customer, its performance may change constantly in different tasks. As for the processors, they are required to submit their feedbacks and trust values after each task. However, some selfish or malicious processors may provide untruthful feedbacks, or some attackers outside the tasks may give fake feedbacks for their own benefits or with the intention of disrupting the entire system. For example, a healthcare center may hire some processors (e.g., patients' wearable devices, surrounding vehicles or smartphones) to help it in providing medical care services. However, a competitor, which may be another healthcare center, may maliciously provide negative comments, hoping that the processors would not join the tasks released by this center.

\subsection{Design Goals}
Using earlier described system model, the goal is to build a trust-based and privacy-preserving customer selection scheme in ubiquitous computing. In particular, the following objectives should be captured.
\begin{itemize}
	\item Privacy: The proposed scheme should preserve processors' privacy. That is, other parties cannot infer processors' location and trajectory information based on the given data.
	\item Security: The proposed scheme should defend against different sophisticated attacks, such as badmouth attack and on-off attack. In addition, some processors may provide fake trust values and feedbacks. The proposed scheme must be resilient to these attacks.
	\item Accuracy: The proposed scheme should accurately calculate the reputation scores of the customers according to processors' feedbacks. Besides, the scheme should identify malicious and honest processors, and further give prediction of their future trust values.
\end{itemize}

\section{Preliminaries}
Bilinear pairing and Paillier cryptosystem are two foundation elements in the proposed scheme. Hence, we introduce them in this section.
\subsection{Bilinear Pairing}
Let $\mathbb{G}$ and $\mathbb{G}_T$ be two multiplicative cyclic groups of the same large prime order $q$. Then, the following three properties can be satisfied by a bilinear map $e: \mathbb{G} \times \mathbb{G} \rightarrow \mathbb{G}_T$.
\begin{itemize}
	\item Bilinear: $e(aP, bQ) = e(P,Q)^{ab}$, for all $P, Q \in \mathbb{G}$ and $a,b \in \mathbb{Z}^*_q$.
	\item Non-degenerated: $e(P,P) \neq 1$, for any $P \in \mathbb{G}$.
	\item Computable: $e(P,Q)$ can be efficiently computed  for all $P,Q \in \mathbb{G}$.
\end{itemize}

We refer to \cite{AbdallaBR01, BonehF01, XuLWZH17} to provide a more comprehensive description and definition for this technique.
\begin{definition}
	A bilinear parameter generator $\mathcal{G}en$ is a probabilistic algorithm which takes a security number $\kappa$ as input, and outputs a 5-tuple $(q, P, \mathbb{G}, \mathbb{G}_T, e)$, where $q$ is a large prime with $\kappa$ bits, $(\mathbb{G}, \mathbb{G}_T)$ are two multiplicative groups with the same order $q$, $P \in \mathbb{G}$ is a generator, and $e: \mathbb{G} \times \mathbb{G} \rightarrow \mathbb{G}_T$ is an efficiently computable bilinear group with the property of non-degeneracy.
\end{definition}

\begin{definition}[Computational Diffie-Hellman (CDH) Problem]
	Given the elements $(P, aP, bP) \in \mathbb{G}$, there exists no probabilistic and polynomial time algorithm to calculate $abP \in \mathbb{G}$ with non-negligible probability of success.
\end{definition}

\subsection{Paillier Cryptosystem}
This cryptosystem is a form of encryption which supports multiplication operations on the ciphertexts. Due to the homomorphic properties, it has been widely used in various privacy-preserving applications \cite{SangST09}. Fundamentally, it consists of the following three algorithms:
\begin{itemize}
	\item {\em Key Generation:} Given a large security parameter $\kappa_1$, and two large primes $p_1, q_1$, where $|p_1| = |q_1| = \kappa_1$. Then, $n = p_1q_1$ and $\lambda = lcm(p_1, q_1)$ are computed, where $lcm(a,b)$ is a function to compute the least common multiple of $a$ and $b$. Define a function $L(c) = \frac{c-1}{n}$, $\mu$ is calculated as $(L(g^\lambda \ \text{mod} \ n^2))^{-1} \ \text{mod} \ n$, where $g \in Z^{*}_{n^2}$ is randomly chosen. Then, the public key $pk$ and secret key $sk$ are generated as $pk=(n,g)$ and $sk=(\lambda,\mu)$ respectively.
	\item {\em Encryption:} Given a message $m \in \mathbb{Z}_n$, the ciphertext is calculated as $c =E(m) = g^m \cdot r^n \ \text{mod} \ n^2$, where $r \in \mathbb{Z}^*_{n}$ is randomly chosen.
	\item {\em Decryption:} Given a ciphertext $c \in \mathbb{Z}^*_{n^2}$,  the ciphertext can be decrypted as $m = D(c) = L(c^{\lambda \ \text{mod} \ n^2}) \cdot \mu \ \text{mod} \ n$. The correctness and security of the Paillier cryptosystem has been proven in \cite{Paillier99}.
\end{itemize}

In particular, the Paillier cryptosystem satisfies the following homomorphic properties:
\begin{itemize}
	\item For any $m_1, m_2 \in \mathbb{Z}_n$, $E(m_1) \cdot E(m_2) = E(m_1 + m_2)$.
	\item For any $m_1, a \in \mathbb{Z}_n$, $E(m_1)^a = E(am_1)$.
\end{itemize}

\section{TPCS Scheme}
The proposed trust-based and privacy preserving customer selection scheme includes system initialization, system overview, report generation, report aggregation, feedback evaluation, and trust value evaluation.

\subsection{System Initialization}
Given security parameters $\kappa$ and $\kappa_1$, TA first generates a 5-tuple $(q, P, \mathbb{G},$ $ \mathbb{G}_T,e)$ by executing $\mathcal{G}en(\kappa)$, and generates the public key $(n=p_1 \cdot q_1,g)$ and private key $(\lambda,\mu)$ of the Paillier cryptosystem. Then, TA selects two secure cryptographic hash functions $H$ and $H_1$, where $H: \{0,1\}^* \rightarrow \mathbb{G}$ and $H_1: \{0,1\}^* \rightarrow \{0,1\}^{\kappa_1} $. Before joining the system, all customers, processors, and RSUs are required to register themselves with TA. Specifically, TA selects a secure symmetric encryption algorithm $AES_{k_0}$ by choosing a symmetric key $k_0$. For every registered customer or processor $v_i$ with its real identity $ID_i$, TA creates a group of pseudonyms $\{PID_{i0}, PID_{i1}, \cdots, PID_{iN}\}$, and generates the public and private key pairs as $Y_{ij} = x_{ij} P$ for $j=\{0,1,\cdots,N\}$, where $x_{ij} \in \mathbb{Z}^*_{q}$ is a random value and $PID_{ij} = AES_{k_0}{(ID_i||x_{ij})}$. Then, TA selects a secure number $\chi \in \mathbb{Z}^*_{n}$ to encrypt each processor's trust value $T_i$ as $C_i = g^{T_i} \cdot (r_i \cdot H_1{(t_c||\chi)})^n \ \text{mod} \ n^2$, where $t_c$ is the current update time and $r_i \in \mathbb{Z}^*_{n}$, and then generates the corresponding trust signature as $\mathfrak{C}_i =C_i \cdot g^{H_1{(t_c||\chi)}} = g^{T_i + H_1{(t_c||\chi)}} \cdot r^n_i \ \text{mod} \ n^2$. Note that the trust signature is used to verify if the trust value is fresh. For every registered RSU, TA selects a random element $x_r \in \mathbb{Z}^*_{q}$ as secret, and calculates the public key as $Y_r = x_r P$. Finally, TA sends the parameter $\{\{PID_{ij}, x_{ij}, Y_{ij}\}^N_{j=1}, t_c, C_i, \mathfrak{C}_i, n, \mathbb{G}, \mathbb{G}_T, e, H\}$ to each customer or processor, $\{n,g, \chi, P, \mathbb{G}, \mathbb{G}_T, e, x_r, Y_r,$ $ H\}$ to each RSU, and $\lambda$ to SP.

\subsection{Scheme Overview}
\begin{figure*}[htb]
	\centering
	\includegraphics[width=0.9\textwidth]{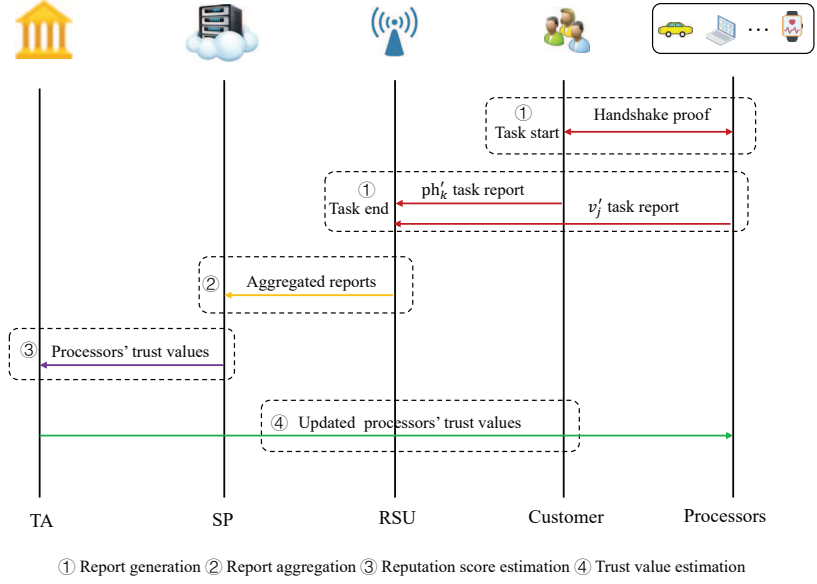}
	\caption{Overview of TPCS.}
	\label{fig:procedure}
\end{figure*}

When a processor joins a task, it first creates a handshake proof with the customer to prove that it has joined this task. After finishing the task, both processor and customer are required to generate their own task reports and deliver them to RSU. Then RSU verifies the processor's validity, i.e., to verify the processor's task report, handshake proof, and trust value. It then uses processors' trust values and feedbacks to calculate the customer's reputation score. Following this, the RSU delivers the reputation score and feedbacks to SP, which will be then used to evaluate the customers' performances. Finally, SP sends processors' trust values to TA, and TA will predict their future performances based on the historical data. This complete process is shown in Fig.~\ref{fig:procedure}.

\subsection{Report Generation}
When a processor $v_j$ with $(\text{PID}_{j}, x_j, C_j,\mathfrak{C}_j)$ finishes a task organized by a customer $\text{ph}_k$, it is required to send a task report to the nearby RSU, which is denoted as $R_j$. Specifically, $v_j$ generates the report, including the customer  $\text{ph}_k$, task $Tr_k$, feedback $f_{j}$, and a handshake proof with $\text{ph}_k$. The handshake proof is used to prove that whether $v_j$ actually joined the $\text{ph}_k$'s task. This proof is generated as follows.
\begin{itemize}
	\item The customer $\text{ph}_k$ generates its Paillier Cryptosystem's public key $(n_k, g_{k})$ and the secret key $(\lambda_k,\mu_k)$. Then, it broadcasts a random value $\alpha_k \in \mathbb{Z}^*_{n_k}$ and its public key to all processors.
	\item The processor $v_j$ selects $\alpha_j, r_j \in \mathbb{Z}^*_{n_k}$ and uses the customer's homomorphic encryption ($n_k$ and $g_k$) to calculate $C_{\alpha_j} = g^{\alpha_j}_k \cdot r^{n_k}_j \ \text{mod} \ n^2_k$, which is the ciphertext of $\alpha_j$. Then, the processor delivers the ciphertext $C_{\alpha_j}$ to the customer.
	\item After receiving the ciphertext, the customer recovers $\alpha_j$ and calculates the proof as $\text{proof}_{kj} = x_k H(\alpha_k + \alpha_j)$. Accordingly, the processor calculates its proof as $\text{proof}_{jk} = x_j H(\alpha_j + \alpha_k)$.
\end{itemize}

Then, to prevent the RSU or other attackers linking $v_j$'s trust value, $v_j$ selects a random vale $r'_j \in \mathbb{Z}^*_{n}$ to perturb the trust ciphertext as $\widetilde{C}_j = g^{T_j} \cdot (r_j \cdot H_1{(t_c||\chi)})^n \cdot (r'_j)^n \ \text{mod} \ n^2$ \footnote{In this paper, $g$ is not public to the processors, and processors can only use the public key $n$ to perturb the ciphertexts.}. Accordingly, the trust signature is also recalculated as $\widetilde{\mathfrak{C}_j} = g^{T_j +  H_1{(t_c||\chi)}} \cdot r^n_j \cdot (r'_j)^n \ \text{mod} \ n^2$. After that, $v_j$ uses $x_j$ to generate a signature as $\sigma_j = x_j(PID_j || Y_j || FR_j || TR_j || $ $ \text{proof}_{jk})$, where $FR_j = (\text{ph}_k||Tr_k||f_j)$ is the feedback report, and $TR_j = (\widetilde{C}_j||\widetilde{\mathfrak{C}_j}||t_c)$ is the trust report. Finally, $v_j$ submits the report $R_j= (PID_j,Y_j,FR_j,TR_j,\text{proof}_{jk},\sigma_j)$ to RSU when it finishes the task. Accordingly, $\text{ph}_k$ uploads its report $R_k = (\text{ph}_k, Tr_k, \{\text{proof}_{kj}\}^{num}_{j=1}, \sigma_k)$, where $num$ is the total number of processors in the task $Tr_k$ and $\sigma_k = x_k H(\text{ph}_k || Tr_k || \{\text{proof}_{kj}\}^{num}_{j=1})$.

\subsection{Report Aggregation}
Upon receiving the reports, RSU first verifies the processor's signature $\sigma_j$, i.e., to check whether $e(P,\sigma_j) \overset{?}{=} e(Y_j,  H(PID_j || Y_j || FR_j || TR_j || \text{proof}_{jk}))$. If it does hold, the signature is valid and RSU will accept $v_j$'s report, since $e(P, \sigma_j) = e(P, x_j H(PID_j || Y_j || FR_j || TR_j || \text{proof}_{jk})) = e(Y_j, H(PID_j || Y_j || FR_j || TR_j || \text{proof}_{jk}))$. To improve verification efficiency with less overhead, RSU can perform batch verification as:
\begin{equation}
\begin{aligned}
e(P, \sum^{sum}_{j=1} \sigma_j) &= e (P, \sum^{sum}_{j=1}x_jH(PID_j || Y_j || FR_j || TR_j || \text{proof}_{jk})) \\
&=\prod^{sum}_{j=1}e(P, x_j H(PID_j || Y_j || FR_j || TR_j || \text{proof}_{jk})) \\
&=\prod^{sum}_{j=1} e (Y_j, H(PID_j || Y_j || FR_j || TR_j || \text{proof}_{jk})).
\label{e1}
\end{aligned}
\end{equation}
By this way, the verification can be completed by executing only $sum +1$ rather than $2 sum$ pairing operations.

After the validity checking, RSU will verify $v_j$'s handshake proof, i.e., $\text{proof}_{jk}$, to check whether it has joined the task. Specifically, RSU verifies $e(Y_k,\text{proof}_{jk}) \overset{?}{=} e(Y_j,\text{proof}_{kj})$. If it holds, the proof is verified, since $e(Y_k, \text{proof}_{jk}) = e(x_k P, x_j H(\alpha_j + \alpha_k)) = e (x_j P, x_k H(\alpha_k+ \alpha_j)) = e(Y_j, \text{proof}_{kj})$. Similarly, RSU can also perform batch verification, that is, to check if $e(Y_k, $ $\sum^{sum}_{j=1} \text{proof}_{jk})  \overset{?}{=} \prod^{sum}_{j=1}e(Y_j, \text{proof}_{kj})$. The proof is given as follows.

\begin{equation}
\begin{aligned}
e(Y_k, \sum^{sum}_{j=1} \text{proof}_{jk}) &= e(Y_k, \sum^{sum}_{j=1}x_j H(\alpha_j + \alpha_k))  \\
&= \prod^{sum}_{j=1} e(x_k P, x_j H(\alpha_j + \alpha_k)) \\
&= \prod^{sum}_{j=1} e (P, x_kx_j H(\alpha_j + \alpha_k))  \\
&= \prod^{sum}_{j=1} e (x_j P, x_k H(\alpha_j + \alpha_k)) \\
&=\prod^{sum}_{j=1} e(Y_j, \text{proof}_{kj}).
\end{aligned}
\end{equation}

Besides the above operations, it is also important to check if the trust value, i.e., $T_j$, is truthful and fresh, as some malicious processors may change their trust values. To achieve this goal, RSU first checks the time stamp $t_c$, and then checks the trust signature $\widetilde{\mathfrak{C}_j}$. Specifically, RSU checks if $\widetilde{C}_j \cdot g^{H_1{(t_c||\chi)}}$ equals to $\widetilde{\mathfrak{C}_j} \cdot {(H_1{(t_c||\chi)})}^{n}$, as $\widetilde{C}_j \cdot g^{H_1{(t_c||\chi)}} = g^{T_j} \cdot (r_j \cdot H_1{(t_c||\chi)})^n \cdot (r'_j)^n \cdot g^{H_1{(t_c||\chi)}} = g^{T_j +  H_1{(t_c||\chi)}} \cdot (r_j \cdot r'_j \cdot H_1{(t_c||\chi)})^n = \widetilde{\mathfrak{C}_j} \cdot {(H_1{(t_c||\chi)})}^{n}$. Similarly, RSU can perform batch verification to check $g^{H_1{(t_c||\chi)}} \cdot \sum^{sum}_{j=1}\widetilde{C}_j \overset{?}{=} {(H_1{(t_c||\chi)})}^{n}  \cdot \sum^{sum}_{j=1} \widetilde{\mathfrak{C}_j}$. The proof is given as follows.
\begin{equation}
\begin{aligned}
g^{H_1{(t_c||\chi)}} \cdot \sum^{sum}_{j=1}\widetilde{C}_j &= \sum^{sum}_{j=1} (g^{T_j + H_1{(t_c||\chi)}} \cdot (r_j \cdot r'_{j} \cdot H_1{(t_c||\chi)})^n) \\
& =  (H_1{(t_c||\chi)})^n \cdot \sum^{sum}_{j=1} \widetilde{\mathfrak{C}_j}
\end{aligned}
\end{equation}

RSU performs the following steps to generate the aggregated report.
\begin{itemize}
	\item Step 1. Compute the aggregated weighted data according to $\{f_{j}, \widetilde{{C}_j}\}^{sum}_{j=1}$ as
	\begin{equation}
	\begin{aligned}
	C_1 &= \prod^{sum}_{j=1} \widetilde{C}^{f_{j}}_j \ \text{mod} \ n^2 \\
	&=\prod^{sum}_{j=1} g^{T_j f_j} \cdot (r_j \cdot r'_j \cdot H_1{(t_c||\chi)})^{n f_j} \ \text{mod} \ n^2 \\
	&= g^{\sum^{sum}_{j=1} T_j f_j} \cdot (\prod^{sum}_{j=1} (r_j r'_j H_1{(t_c||\chi)})^{f_j})^n \ \text{mod} \ n^2 \\
	\end{aligned}
	\end{equation}
	
	\begin{equation}
	\begin{aligned}
	C_2 & = \prod^{sum}_{j=1} \widetilde{C}_j \ \text{mod} \ n^2 \\
	&=g^{\sum^{sum}_{j=1} T_j} \cdot (\prod^{sum}_{j=1} (r_j r'_j H_1{(t_c||\chi)}))^n \ \text{mod} \ n^2  \\
	\end{aligned}
	\end{equation}
	\item Step 2: Use the private key $x_r$ to generate a signature $\sigma_g$ as
	\begin{equation}
	\sigma_r = x_r H(\text{ph}_k || Tr_k || C_1 || C_2 || \{PID_j || f_{j}\}^{sum}_{j=1}).
	\end{equation}
	\item Step 3: Deliver the integrated report $R_r = (\text{ph}_k, Tr_k, C_1, C_2, \{PID_j, f_{j}\}^{sum}_{j=1},$ $ \sigma_r)$ to SP.
\end{itemize}

\subsection{Reputation Score Evaluation}
After receiving the report $R_r$, SP first validates the report by checking if $e(P, \sigma_r)$ equals to $e(Y_r, H(\text{ph}_k || Tr_k || C_1 || C_2 || \{PID_j || f_{j}\}^{sum}_{j=1}))$. Then TA decrypts $C_1, C_2$ by using the secret key $\lambda, \mu$, and calculates the reputation score $RS_{k}$ as follows.
\begin{equation}
\begin{aligned}
RS_k &= \frac{D(C_1)}{D(C_2)}= \frac{L(C_1^{\lambda \ \text{mod} \ n^2}) \cdot \mu \ \text{mod} \ n}{L(C_2^{\lambda \ \text{mod} \ n^2}) \cdot \mu \ \text{mod} \ n} \\
&=\frac{\sum^{sum}_{j=1}T_j f_j}{\sum^{sum}_{j=1}T_j}.
\end{aligned}
\end{equation}

Note that, $RS_k$ is calculated based on processors' previous trust values. To evaluate the qualities of processors' feedbacks in the task $Tr_k$, we design a filtering truth discovery based evaluation algorithm. The basic idea is to assign a higher weight to a processor if its data is closer to the reputation score, and the data provided by a processor with higher weight will be more likely to be considered as the truthful reputation score \cite{LiLGZFH14,LiLGSZFH16,MiaoJSLGQXGR15,MiaoSJLT17,ZhangZXSDG19}. More specifically, $\mathcal{V}_k = [v_1, v_2, \cdots,v_{sum}]$ represents the set of processors which belongs to the task $Tr_k$, and is updated in each iteration since some processors may be removed. The filtering truth discovery based algorithm is achieved by the following steps.
\begin{itemize}
	\item Data filtering: For a processor  $v_j \in \mathcal{V}_k$, SP calculates the difference between each processor's feedback and the reputation score, and then removes the processors whose difference is less than a threshold, i.e.,
	\begin{equation}
	\label{td_r}
	|f_j - RS^{(v)}_k| < U_{threshold},
	\end{equation}
	where $RS^{(v)}_k$ denotes $\text{ph}_k$'s reputation score in the $v$-th iteration.
	
	\item Weight update: SP calculates the difference between each processor's feedback and the customer's reputation score, and then updates each processor's weight based on the aggregated differences. Without loss of generality, we adopt a logarithmic weight function, which has been widely used in truth discovery based applications \cite{MiaoJSLGQXGR15,ZhangZXSDG19}.
	\begin{equation}
	\label{td_w}
	w^{(v)}_{j} = \log(\frac{\sum_{v_j \in \mathcal{V}^{(v)}_k} d(f_{j}, RS^{(v)}_{k})}{d(f_{j},RS^{(v)}_k)}),
	\end{equation}
	where $d(\cdot)$ is a distance function calculated as $d(f_{j},RS^{(v)}_k) = (f_{j} - RS^{(v)}_k)^2$.
	\item Reputation score update: Based on the processors' weights, the reputation score for the customer can be estimated as
	\begin{equation}
	\label{td_t}
	RS^{(v+1)}_{k} = \frac{\sum_{v_j \in \mathcal{V}^{(v)}_{k}}w^{(v)}_{j} \cdot f_{j}}{ \sum_{v_j \in  \mathcal{V}^{(v)}_{k}} w^{(v)}_{j}}.
	\end{equation}
\end{itemize}

The above procedures will be iteratively conducted until the change of the reputation score between two consecutive iterations is less than a predefined threshold. Then, SP publishes the customer's reputation score. The general procedure of the filtering truth discovery based evaluation algorithm is shown in Algorithm \ref{A1}.

\begin{algorithm}[!htbp]
	\label{A1}
	\SetCommentSty{small}
	\LinesNumbered
	\caption{Filtering truth discovery based evaluation algorithm}
	\KwIn{processors' feedbacks $\{f_j\}^{sum}_{j=1}$}
	\KwOut{Reputation score $RS_k$, processors' trust values $\{T_j\}^{sum}_{j=1}$}
	\For{$iteration=1,2,\cdots,v$}{
		Update the set of processors (see Eq.~\ref{td_r});  \\
		\For{$v_j \in \mathcal{V}^{(v)}_{k}$}{
			Update processors' trust values (see Eq.~\ref{td_w})\;}
		{
			Update the reputation score (see, Eq.~\ref{td_t})};
	}
	\Return{$RS_k$;}
\end{algorithm}

\subsection{Trust Value Evaluation}
Based on the reputation score, SP can also obtain processors' new trust values. Motivated by \cite{HuLZS17}, we define a function to measure the qualities of processors' trust values.
\begin{equation}
\label{e13}
T_j =\left\{
\begin{array}{rcl}
1-|f_j - RS_k|^{v \cdot c_0}      &      &  |f_j - RS_k | < F_{threshold}\\
1-|f_j - RS_k|   &      & otherwise\\
\end{array} \right.
\end{equation}
where $v$ is the number of iterations. It is obvious that if there are more malicious processors, $v$ will be larger and it will be more difficult to obtain the accurate reputation score. Thus, $v$ can be used as a reward for the processors whose feedbacks contribute to the accurate reputation score calculation. Besides, we define another factor $c_0$ to control the reward sensitivity. If the difference between a processor's feedback and the reputation score is more than a threshold $F_{threshold}$, then the feedback does not make any positive effect on the reputation score and hence the processor will not obtain the reward. Then, SP delivers processors' trust values  $[(PID_1, T_1), (PID_2, T_2), \cdots, (PID_{sum},T_{sum})]$ to TA.

On receiving the trust values, TA first uses the symmetric key $t_0$ to retrieve processors' real identities, and then predicts processors' future trust values according to their historical behaviors. Here, we use the exponential weighted moving average (EMWA) technique to estimate processors' future behaviors, as it gives more consideration of processors' most recent performances \cite{Vendramin2012GrAnt,XiaLLAYM15}.
\begin{equation}
T_{j(i+1)} = \alpha \times T_{j(i-1)} + (1 - \alpha) \times T_{ji},
\end{equation}
where $\alpha \in (0,1)$ is an impact factor, and $T_{j(i-1)}, T_{ji}$ and $T_{j(i+1)}$ are the past, current and future trust values respectively .

Note that, some processors may behave well at the beginning to improve their trust values, and behave badly when these values are high enough. To counter the effect of this attack, we further design a trust value circuit-breaker mechanism as:
\begin{equation}
\label{e13}
T_{j(i+1)}=\left\{
\begin{array}{rcl}
T_0      &      &  T_{j(i-1)} - T_{ji} > T_{threshold}\\
T_{j(i+1)}   &      & otherwise\\
\end{array} \right.
\end{equation}
From this equation, we can see if the decrease between two consecutive trust values is larger than a predefined threshold, the trust value will be set as the initialized value $T_0$. Moreover, to punish the on-off attacker, once the circuit-breaker is triggered, the predicted trust value will be decreased as $T_{j(i+1)} = c_1 \cdot T_{j(i+1)}$, where $c_1 \in (0,1)$ is a forgetting factor. In this way, the attacker will take more time to bring its trust value to the previous level.

\section{Security Analysis}
Before presenting the evaluation, we first discuss the security analysis of our proposed TPCS scheme. In particular, according to the security model discussed earlier, we first focus on how TPCS scheme can achieve processor's report privacy preservation, authentication and data integrity, and then we discuss some attack strategies and give the resilience analysis against them.

\subsection{Security analysis for processor's report}
{\em The processor's report is privacy-preserving.} In this proposed scheme, processor's trust information is encrypted as a valid Paillier ciphertext $C_j = g^{M_j} \cdot R^n_j \ \text{mod} \ n^2$ if we consider the trust values $T_j$ and $T_j + H_1{(t_c||\chi)}$ as the message $M_j$, and the random values $r_j$, $r_j \cdot H_1{(t_c||\chi)}$, and $r_j \cdot H_1{(t_c||\chi)} \cdot r'_j$ as $R_j$. As the Paillier Cryptosystem can defend from the chosen plaintext attack, the trust value achieves semantic security and privacy preservation. Hence, although an adversary may eavesdrop the ciphertext $C_j$, it cannot identify the original data. After collecting processors' reports, RSU will compute $C_1$ and $C_2$ to aggregate all reports. However, the RSU or an adversary cannot get each individual's trust value without the secret key. Finally, SP can recover $C_1$ and $C_2$ as $\sum^{sum}_{j=1} T_j \cdot f_j$ and $\sum^{sum}_{j=1} T_j$. Nevertheless, since the decrypted data is aggregated results, it cannot get each individual processor's trust value $(T_1, T_2, \cdots, T_{sum})$. Therefore, the processor's report is privacy-preserving.

{\em The processor's report achieves authentication and data integrity.} The processor's report is signed using the BLS short signature \cite{BonehLS04}. As the BLS signature has been proven to be secure under the CDH problem \cite{BellareR93}, any malicious behaviors of an adversary can be detected, and accordingly our proposed scheme can guarantee the report's authentication and data integrity.

\subsection{Resilience analysis against attacks launched by adversaries}

{\em Resilience to link attack:} From TPCS scheme's perspective, link attack means that an attacker may link a certain  $v_j$ to its identity or trust value. To prevent the identity link attack, $v_j$ can change its pseudonym when it joins different tasks, which will make them unlinkable. However, given that the trust value remains unchanged for some time, it may still be linked according to its trust value. In our proposed scheme, $v_j$ does not submit its original trust values directly to the RSU. Instead, the trust value is encrypted, and $v_j$ changes the ciphertext by multiplying a random value $ (r'_j)^n$ when it takes part in a different task  (i.e., $C_j \rightarrow \widetilde{{C}_j} = C_j \cdot (r'_j)^n \ \text{mod} \ n^2$). Besides, although SP owns the secret key $\lambda$, it still cannot trace any individual processor as the encrypted trust values have been aggregated in RSU before uploading to SP.

{\em Resilience to fake trust value attack:} In this scheme, the trust value is encrypted and hence the processor  has no method to know its real trust level. However, some malicious processor may provide a fake trust value by colluding with other processors or using its previous data. In such a case, our scheme is still effective as we use time stamp $t_c$ to generate the trust ciphertext and trust signature, i.e., ($C_j = g^{T_j} \cdot (r_j \cdot H_1{(t_c||\chi)})^n \ vs. \ \mathfrak{C}_j = g^{T_j + H_1{(t_c||\chi)}} \cdot r^n_j$ ). Specifically, for the first collusion attack ($v_j$ is in collusion with $v_i$ for example), $v_j$ submits its falsified trust report as $\widehat{C}_j = C_j \cdot C_i = g^{T_j + T_i} \cdot (r_j r_i (H_1{(tc || \chi}))^2)^n$, $\widehat{\mathfrak{C}_j} = \mathfrak{C}_j \cdot \mathfrak{C}_i = g^{T_j + T_i + 2H_1{(t_c||\chi)}} \cdot (r_j r_i)^n$. Since $\widehat{C}_j \cdot g^{H_1{(t_c||\chi)}} \neq \widehat{\mathfrak{C}_j} \cdot {(H_1{(t_c||\chi)})}^{n}$, this malicious manipulation will be identified by RSU. For the second reply attack, $v_j$ submits its previous trust report $C'_j = g^{T'_j} \cdot (r'_j H_1{(t'_c || \chi)})^n$, $\mathfrak{C}'_j = g^{T'_j + H_1{(t'_c || \chi)}} \cdot (r'_j)^n$. Also, it still cannot pass the authentication, as $C'_j \cdot g^{H_1{(t_c||\chi)}} \neq \mathfrak{C}'_j \cdot {(H_1{(t_c||\chi)})}^{n}$.

{\em Resilience to badmouth attack:} From TPCS scheme's perspective, a badmouth attack means that the attackers may always provide low feedbacks for customers. Specifically, the badmouth attackers can be categorized into internal and external attackers. For the external badmouth attackers, we design report authentication and handshake protocols to ensure that only the valid processors which register with the system and join the task can pass the authentication. For the internal badmouth attackers, the proposed scheme incorporates a filtering truth discovery based reputation evaluation algorithm to distinguish among truths which deviate from ground truth.

{\em Resilience to on-off attack:} In the proposed scheme, an on-off attack means that some malicious processors may perform well to hide themselves before launching attacks. When they attain high trust values, they launch attacks and then remain dormant for a certain time period to regain their trust. This attack is hard to defend against using the traditional methods. In our scheme, we design a circuit-breaker mechanism to handle this problem, which is motivated by a common human nature that people make great efforts to build up trust values and some bad behaviors will destroy them. Specifically, we define a $T_{threshold}$, and once the decrease of two consecutive trust values is larger than the predefined threshold, the circuit-breaker will be triggered. Besides, to punish the on-off attacker, its trust value will be decreased by multiplying a forgetting factor $c_0 \in (0,1)$ in a long time. That is, the attacker will take a long time to build up its trust value to the previous level. Thus, our proposed scheme mitigates the on-off attack.

\section{Performance Evaluation}
Here, we evaluate the performance of TPCS scheme in terms of efficiency and effectiveness in task selection. The proposed scheme is implemented in Java, and all experiments are conducted on a system with Intel Core i7 2.5 GHz processor and 16GB RAM. The detailed parameter setting is shown in Table~\ref{tab:1}.

\begin{table}[!h]
	\caption{The Parameters for evaluation}
	\centering
	\label{tab:1}
	
	\begin{tabular}{lll}
		\hline\noalign{\smallskip}
		Notation & Definition & Value  \\
		\noalign{\smallskip}\hline\noalign{\smallskip}
		$\kappa, \kappa_1$ & security parameter & $\kappa, \kappa_1 = 512$ \\
		$q$ &Generator of Bilinear Pairing& $q=512$ \\
		$p_1,p_2$ &Generator of Paillier cryptosystem & $p_1, q_1 = 512$ \\
		$m_h$ & number of customers & $10$ \\
		$sum$ & number of processors & $50$ \\
		$\rho$ & malicious processor proportion & $20\%$ \\
		$T_0$ & initial trust value &$0.01$ \\
		$c_0$ & reward sensitivity & 0.1 \\
		$c_1$ & forgetting factor & 0.85 \\
		$\alpha$ & impact factor parameter & $0.3$ \\
		$v$ & number of iterations& 10 \\
		$U_{threshold}$ & threshold which triggers set update & 0.5 \\
		$F_{threshold}$ &threshold which triggers reward & 0.2 \\
		$T_{threshold}$ &threshold which triggers circuit-breaker &$0.5$ \\
		\noalign{\smallskip}\hline
	\end{tabular}
\end{table}
\subsection{Efficiency Analysis}
\begin{figure*}[htb]
	\centering
	\subfigure[Computational cost for report authentication]
	{ \includegraphics[width=.3\textwidth]{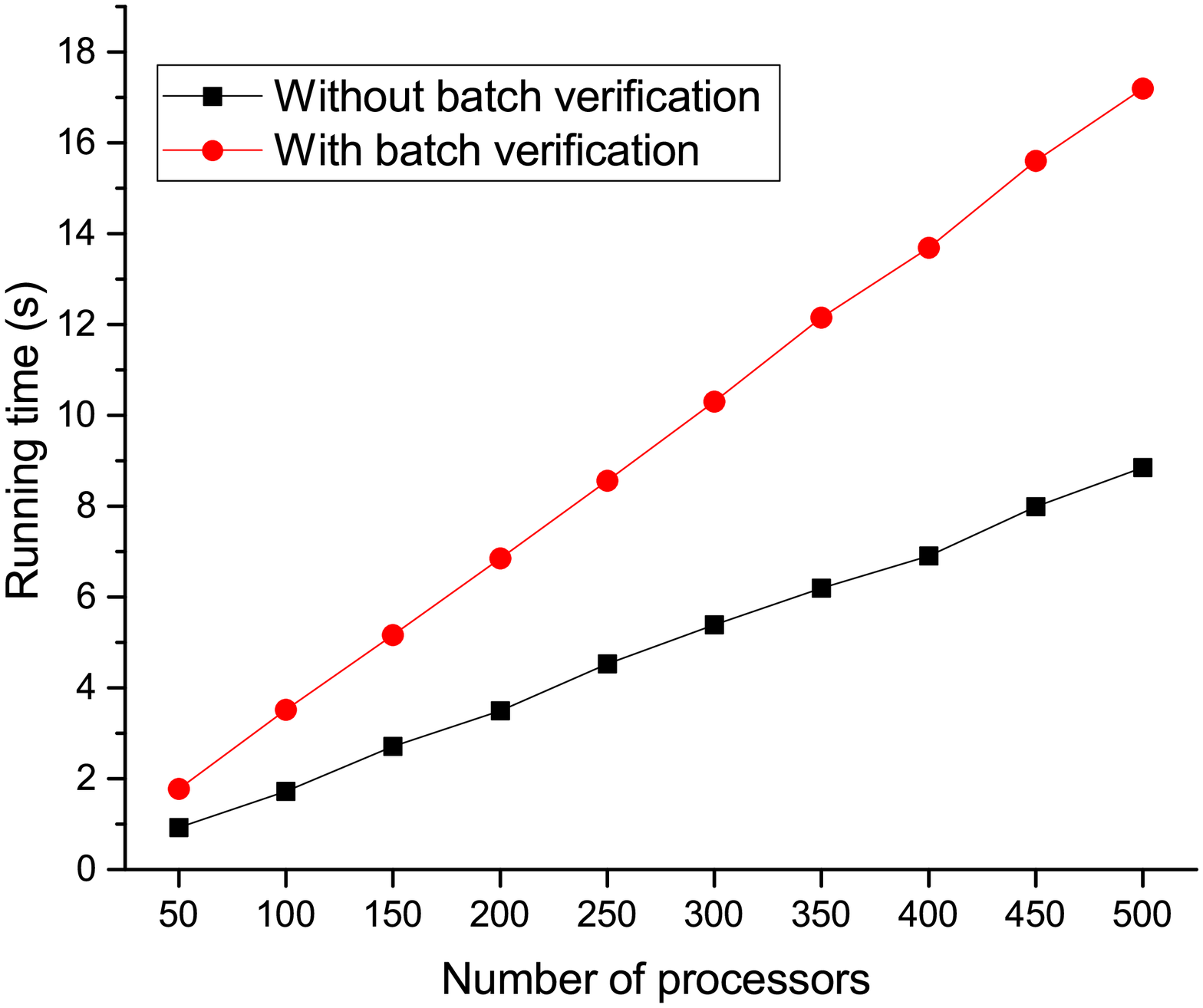}}
	~~
	\subfigure[Computational cost for handshake authentication]
	{\includegraphics[width=.3\textwidth]{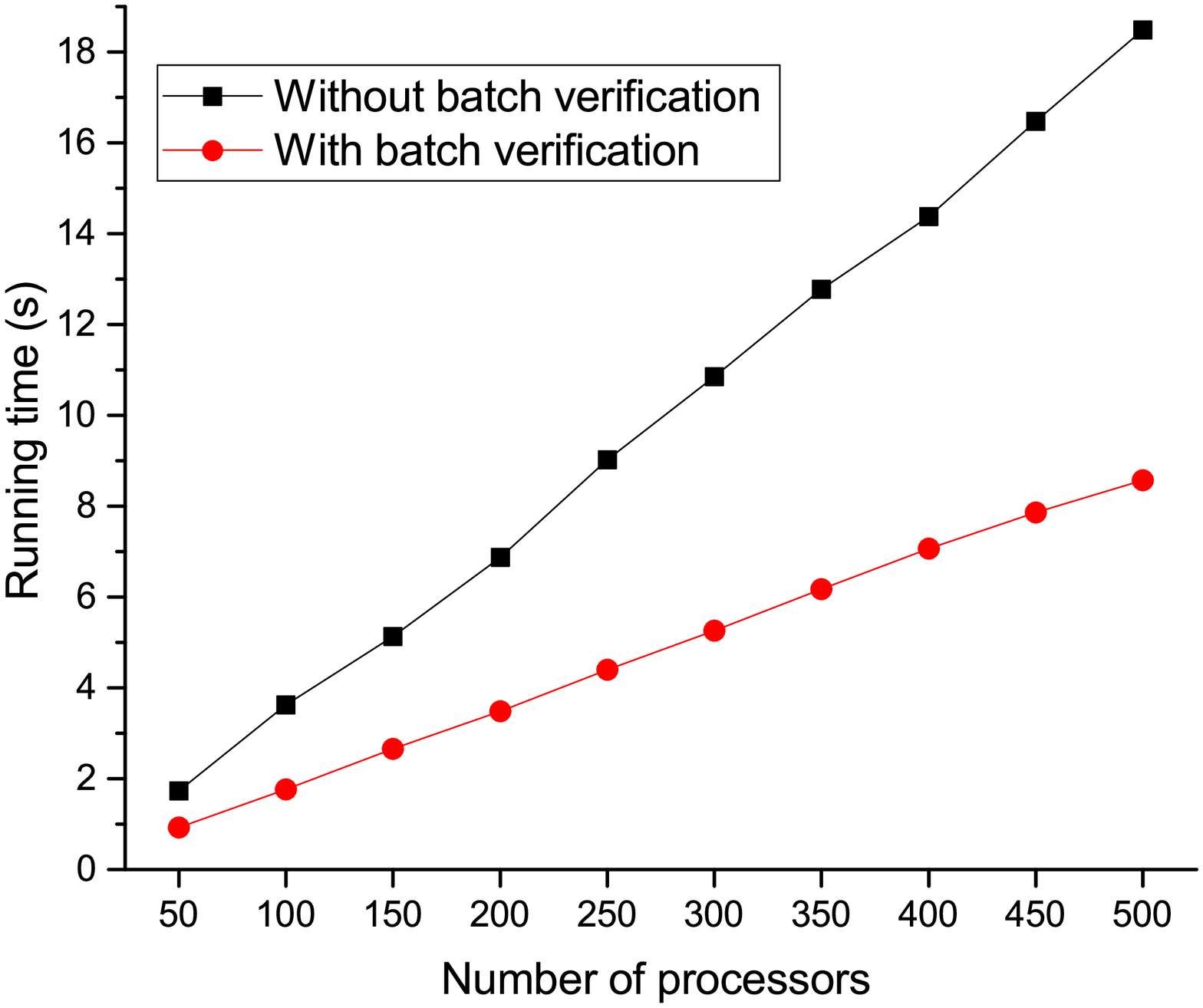}}
	~~
	\subfigure[Computational cost for trust authentication]
	{\includegraphics[width=.3\textwidth]{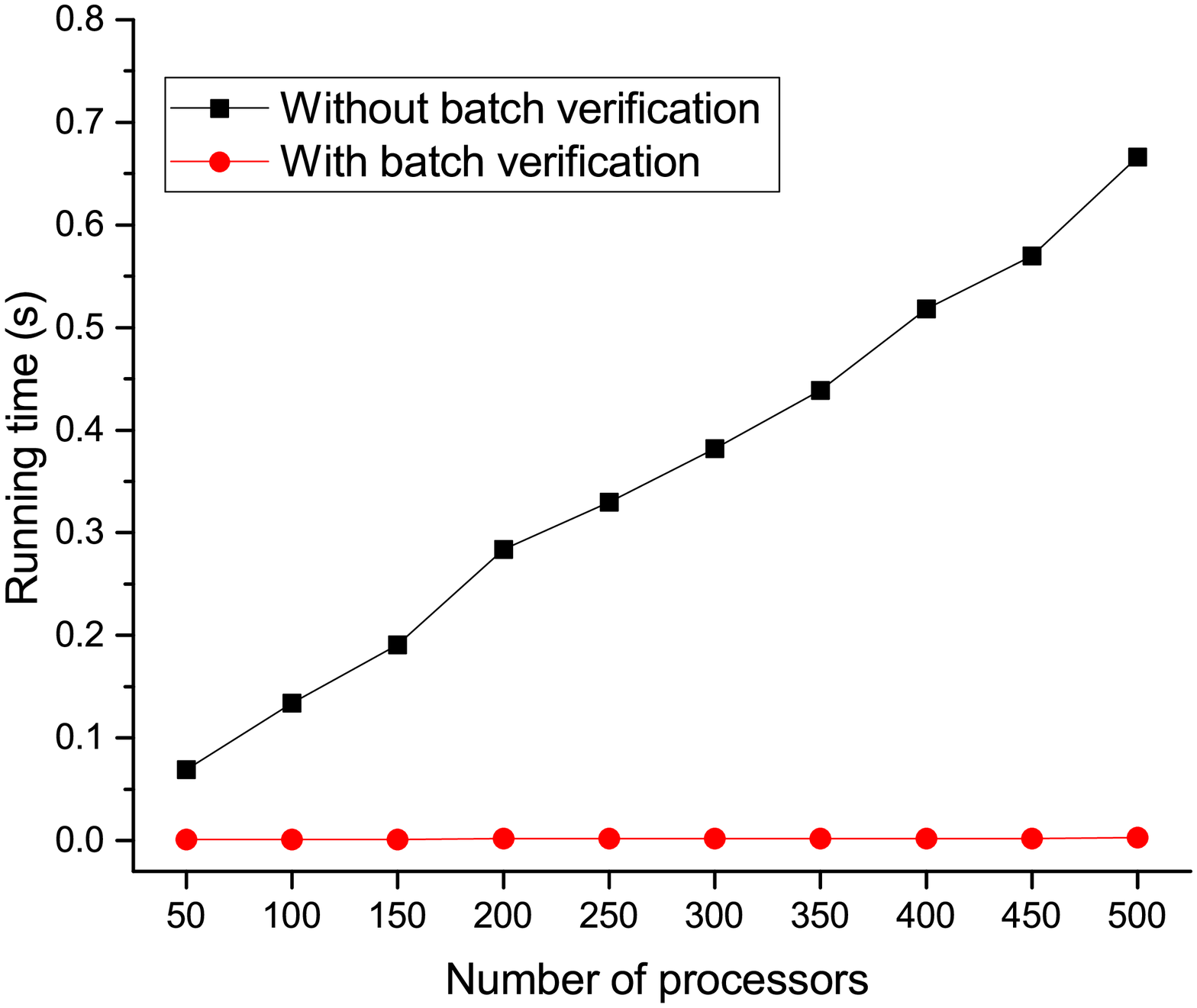}}
	~~
	\caption{Computational cost for each authentication.}
	\label{fig:identification}
\end{figure*}

\begin{figure}[htb]
	\centering
	\includegraphics[ width=8.7cm]{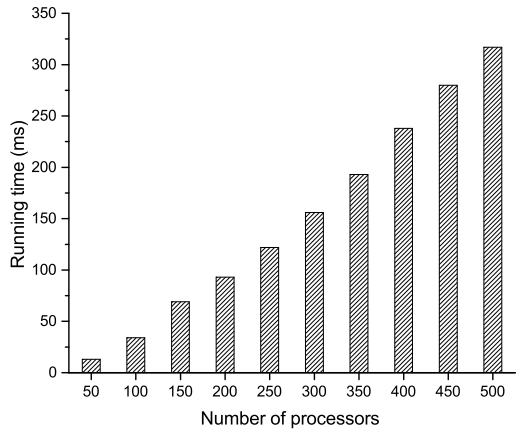}
	\caption{Computational cost for ciphertexts aggregation.}
	\label{fig:aggregation}
\end{figure}

In this experiment, the aim is to evaluate the efficiency of TPCS scheme in terms of authentication and ciphertexts aggregation. Every experiment is executed 10 times and the average result is used for analysis. Note that, in the proposed scheme, three authentication protocols (i.e., report authentication, handshake authentication, trust authentication) are designed for processors' verification. Fig. \ref{fig:identification}  illustrates the computational cost of the authentication varying against the number of processors. As can be seen, since we use batch verification in each authentication, the verification is finished with fewer pairing operations, and accordingly the running time is much less as compared to no batch verification. In Fig. \ref{fig:aggregation}, we plot the running time of ciphertexts aggregation. From this figure, we can observe that as the number of processors increases, our scheme can efficiently perform the ciphertexts aggregation. This is evident from the fact that only 317 ms is required to execute the ciphertext aggregation for 500 processors.

\subsection{Effectiveness Analysis}
In order to analyze the correctness of the system, we vary the percentage of malicious processors. The malicious processors will provide untruthful feedbacks, that is, their feedbacks are much higher or lower than the truthful evaluation. All processors are initialized with the same trust value $T_0$. After the execution of our proposed truth discovery based evaluation algorithm, we observe the value change of the customer  and processors.

\begin{figure}[htb]
	\centering
	\includegraphics[ width=8.7cm]{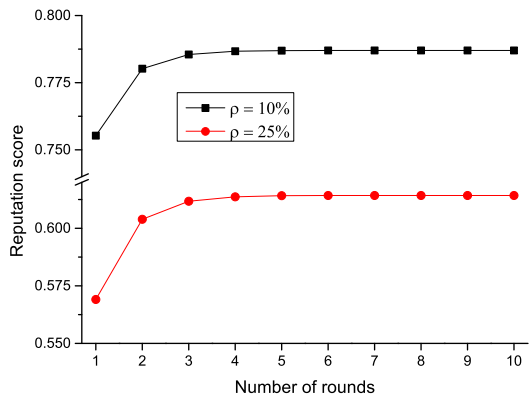}
	\caption{Reputation score comparison ($\rho = 10\%$ and $\rho = 25\%$).}
	\label{fig:reputation_com}
\end{figure}

\begin{figure}[htb]
	\centering
	\includegraphics[ width=8.7cm]{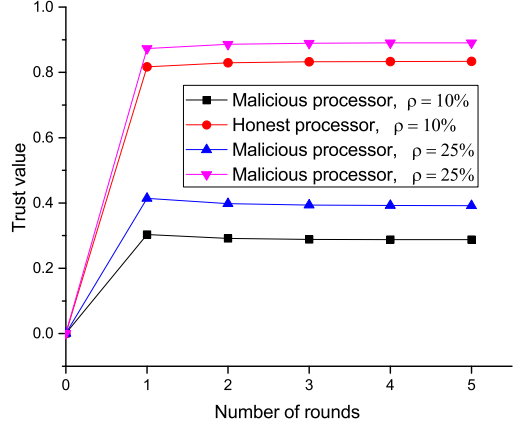}
	\caption{Trust values comparison for honest and malicious processor.}
	\label{fig:trustCom}
\end{figure}

Fig. \ref{fig:reputation_com} plots the reputation score of the customer where the percentage of malicious processors is set as 10 and 25 percent respectively, i.e., $\rho = 10\%$ and $\rho=25\%$. As can be seen, the reputation score tends to be stable after the fourth round. When the number of malicious processors accounts for 10\% of the total number of processors, the reputation score is equivalent to 0.787. When there are more malicious feedbacks, the reputation score witnesses a downward trend, while is still in a reasonable range. Fig. \ref{fig:trustCom} presents the trust value of a single malicious processor and a single honest processor  where $\rho = 10\%$ and $\rho=25\%$. It is obvious that the malicious processor gets the lower trust value after the experiments, which demonstrates the correctness of TPCS scheme.

\begin{figure}[!h]
	\centering
	\includegraphics[ width=8.7cm]{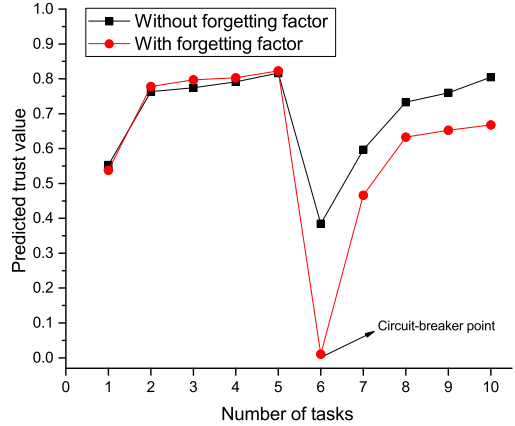}
	\caption{Trust value comparison based on on-off attacks.}
	\label{fig:forgetting}
\end{figure}

We also analyze the effectiveness of the scheme for resilience to on-off attack. Recall that an on-off attack means that a processor behaves well to accumulate its trust value at the beginning, and give untruthful feedback when its trust value is high enough. To mitigate the effect of this attack, we design a circuit-breaker mechanism and apply the forgetting factor. As shown in Fig. \ref{fig:forgetting}, without the forgetting factor (i.e., the black line), the processor  performs well in the first five  tasks, and its trust value rises up to 0.8165. After launching the badmouth attack, its trust value decreases quickly, which however rises up to 0.8045 after only four more  tasks. In contrast, with the forgetting factor (i.e., red line), it triggers the circuit-breaker at the sixth task, and its trust value rises slowly at the later  task. That is, the attacker will need more time to bring its trust value to the previous level and hence demonstrates the effectiveness of TPCS scheme.

\section{Related work}
By taking advantage of the increasingly powerful computation and communication capabilities of smart devices, ubiquitous computing has received considerable attention in recent years \cite{AbowdM00,Lai2013Multi,MaZCCYHJ06}. As a special computing paradigm, ubiquitous computing integrates nearby devices and provides tremendous novel services by exploiting their idle computing and communication resources. This enables complex computing possible at anytime, anywhere, and by different means. Recently, many ubiquitous computing based applications have been studied and proposed, such as data mining \cite{ZhangWW18}, human activity recognition \cite{LiuPWLH16}, disease support \cite{Nieto2017Classification}, and ambient intelligence \cite{Acampora2008A}. However, the selection of a reliable customer while not disclosing processors' privacy is still an unsolved problem.

To the best of our knowledge, none of the existing schemes have solved this problem. Generally, to find a reliable customer and help processors avoid selecting the misbehaving customers, a potential approach is to evaluate the trustworthiness of the customers and processors. Recently, many trust models have been proposed in VANET  \cite{Chen2010A},\cite{HuLZS17},\cite{JavedZH18},\cite{PatwardhanJFY06},\cite{RayaPGH08},\cite{ZhuZXS18}. Specifically, in \cite{PatwardhanJFY06}, Patwardhan {\em et al.} proposed a context-aware reputation management approach for vehicular ad hoc networks, which provides a bootstrapping method to enable vehicles to establish trust relationships. Nevertheless, since it lacks of robustness and scalability, their proposed model cannot be directly used in the task. Raya {et al.} \cite{RayaPGH08} presented a data-oriented trust establishment scheme. However, their framework is not efficient in handling the large amount of feedback data in a task scenario.  Zhu {\em et al.} \cite{ZhuZXS18} described a trust management scheme for vehicular crowdsensing applications. Hu {\em et al.} \cite{HuLZS17} presented a reliable and trust-based task service selection scheme. By building a trust-based evaluation model, their scheme can defend against several sophisticated attacks in VANET. However, their scheme ignores the outside attackers and cannot be executed with an untrusted service provider. Javed \cite{JavedZH18} presented a security adaptation scheme to improve the quality of service (QoS) of safety applications. In their scheme, several factors such as connectivity duration, near s' centrality metrics, and security level are combined to calculate the trust level. However, their work focuses on the QoS in Vehicular Sensor Networks (VSNs) and it is not suitable in the applications of ubiquitous computing. Zhu et al. \cite{ZhuWGC19} used blockchain to realize controllable and trustworthy data management in cloud environment. As for the security and trust in ubiquitous computing, some works have been proposed recently. In \cite{XingJ13}, Xing et al. defined individuals' contexts to be private information and based on this they proposed a context aggregation and sharing scheme which could achieve trust and privacy preservation. In \cite{Iltaf2012Modeling}, an adaptive trust and selection scheme is proposed to realize access control for pervasive environment. In \cite{AkramCLSY18}, Akram te al. summarized the literature related to security, privacy, and trust issues in user-centric solutions. However, to our best knowledge, none of the existing works have focused on how to select reliable customers for the data processors, which inspires to design a trust-based and privacy-preserving customer selection scheme for ubiquitous computing. Besides the above works, we also study the potential security and privacy issues in internet of things and sensor networks \cite{DuC08,DuGXC09,DuXGC07,XiaoRSDHG07,Guan,0001DZHG07}, as the core of ubiquitous computing is to take advantage of various computing devices.

\section{Conclusion}
In this article we proposed a trust-based and privacy-preserving customer selection scheme for processors before they join a ubiquitous computing task. Considering the uncertainty of processors' behaviors, we design a filtering truth discovery based evaluation mechanism to calculate the customers' reputation scores and processors' trust values. In addition, three authentication protocols are designed to ensure that only the valid processors can pass the authentication. Security analysis and simulation results establish the security and effectiveness of the proposed scheme. Using similar approach, we aim to address other trust and privacy challenges in different ubiquitous computing based application scenarios.

\section*{References}
\bibliographystyle{plain}

\bibliography{main}

\begin{thebibliography}{48}
\expandafter\ifx\csname natexlab\endcsname\relax\def\natexlab#1{#1}\fi
\providecommand{\url}[1]{\texttt{#1}}
\providecommand{\href}[2]{#2}
\providecommand{\path}[1]{#1}
\providecommand{\DOIprefix}{doi:}
\providecommand{\ArXivprefix}{arXiv:}
\providecommand{\URLprefix}{URL: }
\providecommand{\Pubmedprefix}{pmid:}
\providecommand{\doi}[1]{\href{http://dx.doi.org/#1}{\path{#1}}}
\providecommand{\Pubmed}[1]{\href{pmid:#1}{\path{#1}}}
\providecommand{\bibinfo}[2]{#2}
\ifx\xfnm\relax \def\xfnm[#1]{\unskip,\space#1}\fi
\bibitem[{Abdalla et~al.(2001)Abdalla, Bellare and Rogaway}]{AbdallaBR01}
\bibinfo{author}{Abdalla, M.}, \bibinfo{author}{Bellare, M.},
  \bibinfo{author}{Rogaway, P.}, \bibinfo{year}{2001}.
\newblock \bibinfo{title}{The oracle diffie-hellman assumptions and an analysis
  of {DHIES}}, in: \bibinfo{booktitle}{Proc. of RSA Conf. on Topics in
  Cryptology}, pp. \bibinfo{pages}{143--158}.
\bibitem[{Abowd and Mynatt(2000)}]{AbowdM00}
\bibinfo{author}{Abowd, G.D.}, \bibinfo{author}{Mynatt, E.D.},
  \bibinfo{year}{2000}.
\newblock \bibinfo{title}{Charting past, present, and future research in
  ubiquitous computing}.
\newblock \bibinfo{journal}{{ACM} Trans. Comput.-Hum. Interact.}
  \bibinfo{volume}{7}, \bibinfo{pages}{29--58}.
\bibitem[{Acampora and Loia(2008)}]{Acampora2008A}
\bibinfo{author}{Acampora, G.}, \bibinfo{author}{Loia, V.},
  \bibinfo{year}{2008}.
\newblock \bibinfo{title}{A proposal of ubiquitous fuzzy computing for ambient
  intelligence}.
\newblock \bibinfo{journal}{Information Sciences} \bibinfo{volume}{178},
  \bibinfo{pages}{631--646}.
\bibitem[{Akram et~al.(2018)Akram, Chen, L{\'{o}}pez, Sauveron and
  Yang}]{AkramCLSY18}
\bibinfo{author}{Akram, R.N.}, \bibinfo{author}{Chen, H.},
  \bibinfo{author}{L{\'{o}}pez, J.}, \bibinfo{author}{Sauveron, D.},
  \bibinfo{author}{Yang, L.T.}, \bibinfo{year}{2018}.
\newblock \bibinfo{title}{Security, privacy and trust of user-centric
  solutions}.
\newblock \bibinfo{journal}{Future Generation Comp. Syst.}
  \bibinfo{volume}{80}, \bibinfo{pages}{417--420}.
\bibitem[{Bellare and Rogaway(1993)}]{BellareR93}
\bibinfo{author}{Bellare, M.}, \bibinfo{author}{Rogaway, P.},
  \bibinfo{year}{1993}.
\newblock \bibinfo{title}{Random oracles are practical: {A} paradigm for
  designing efficient protocols}, in: \bibinfo{booktitle}{Proc. of ACM
  Conference on Computer and Communications Security}, pp.
  \bibinfo{pages}{62--73}.
\bibitem[{Boneh and Franklin(2001)}]{BonehF01}
\bibinfo{author}{Boneh, D.}, \bibinfo{author}{Franklin, M.K.},
  \bibinfo{year}{2001}.
\newblock \bibinfo{title}{Identity-based encryption from the weil pairing}, in:
  \bibinfo{booktitle}{Proceedings of Intl. Conf. on Advances in Cryptology},
  pp. \bibinfo{pages}{213--229}.
\bibitem[{Boneh et~al.(2004)Boneh, Lynn and Shacham}]{BonehLS04}
\bibinfo{author}{Boneh, D.}, \bibinfo{author}{Lynn, B.},
  \bibinfo{author}{Shacham, H.}, \bibinfo{year}{2004}.
\newblock \bibinfo{title}{Short signatures from the weil pairing}.
\newblock \bibinfo{journal}{J. Cryptology} \bibinfo{volume}{17},
  \bibinfo{pages}{297--319}.
\bibitem[{Chen et~al.(2010)Chen, Zhang, Cohen and Ho}]{Chen2010A}
\bibinfo{author}{Chen, C.}, \bibinfo{author}{Zhang, J.},
  \bibinfo{author}{Cohen, R.}, \bibinfo{author}{Ho, P.H.},
  \bibinfo{year}{2010}.
\newblock \bibinfo{title}{A trust modeling framework for message propagation
  and evaluation in vanets}, in: \bibinfo{booktitle}{Proc. of Int. Conf. on
  Information Technology Convergence and Services}, pp. \bibinfo{pages}{1--8}.
\bibitem[{Du and Chen(2008)}]{DuC08}
\bibinfo{author}{Du, X.}, \bibinfo{author}{Chen, H.}, \bibinfo{year}{2008}.
\newblock \bibinfo{title}{Security in wireless sensor networks}.
\newblock \bibinfo{journal}{{IEEE} Wireless Commun.} \bibinfo{volume}{15},
  \bibinfo{pages}{60--66}.
\bibitem[{Du et~al.(2009)Du, Guizani, Xiao and Chen}]{DuGXC09}
\bibinfo{author}{Du, X.}, \bibinfo{author}{Guizani, M.}, \bibinfo{author}{Xiao,
  Y.}, \bibinfo{author}{Chen, H.}, \bibinfo{year}{2009}.
\newblock \bibinfo{title}{Transactions papers a routing-driven elliptic curve
  cryptography based key management scheme for heterogeneous sensor networks}.
\newblock \bibinfo{journal}{{IEEE} Trans. Wireless Communications}
  \bibinfo{volume}{8}, \bibinfo{pages}{1223--1229}.
\bibitem[{Du et~al.(2007)Du, Xiao, Guizani and Chen}]{DuXGC07}
\bibinfo{author}{Du, X.}, \bibinfo{author}{Xiao, Y.}, \bibinfo{author}{Guizani,
  M.}, \bibinfo{author}{Chen, H.}, \bibinfo{year}{2007}.
\newblock \bibinfo{title}{An effective key management scheme for heterogeneous
  sensor networks}.
\newblock \bibinfo{journal}{Ad Hoc Networks} \bibinfo{volume}{5},
  \bibinfo{pages}{24--34}.
\bibitem[{Hei and Du(2011)}]{HeiD11}
\bibinfo{author}{Hei, X.}, \bibinfo{author}{Du, X.}, \bibinfo{year}{2011}.
\newblock \bibinfo{title}{Biometric-based two-level secure access control for
  implantable medical devices during emergencies}, in:
  \bibinfo{booktitle}{{INFOCOM} 2011. 30th {IEEE} International Conference on
  Computer Communications, Joint Conference of the {IEEE} Computer and
  Communications Societies, 10-15 April 2011, Shanghai, China}, pp.
  \bibinfo{pages}{346--350}.
\bibitem[{Hei et~al.(2010)Hei, Du, Wu and Hu}]{HeiDWH10}
\bibinfo{author}{Hei, X.}, \bibinfo{author}{Du, X.}, \bibinfo{author}{Wu, J.},
  \bibinfo{author}{Hu, F.}, \bibinfo{year}{2010}.
\newblock \bibinfo{title}{Defending resource depletion attacks on implantable
  medical devices}, in: \bibinfo{booktitle}{Proceedings of the Global
  Communications Conference, 2010. {GLOBECOM} 2010, 6-10 December 2010, Miami,
  Florida, {USA}}, pp. \bibinfo{pages}{1--5}.
\bibitem[{Hu et~al.(2017a)Hu, Lu and Zhang}]{HuLZ17}
\bibinfo{author}{Hu, H.}, \bibinfo{author}{Lu, R.}, \bibinfo{author}{Zhang,
  Z.}, \bibinfo{year}{2017}a.
\newblock \bibinfo{title}{{TPSQ:} trust-based platoon service query via
  vehicular communications}.
\newblock \bibinfo{journal}{Peer-to-Peer Networking and Applications}
  \bibinfo{volume}{10}, \bibinfo{pages}{262--277}.
\bibitem[{Hu et~al.(2017b)Hu, Lu, Zhang and Shao}]{HuLZS17}
\bibinfo{author}{Hu, H.}, \bibinfo{author}{Lu, R.}, \bibinfo{author}{Zhang,
  Z.}, \bibinfo{author}{Shao, J.}, \bibinfo{year}{2017}b.
\newblock \bibinfo{title}{{REPLACE:} {A} reliable trust-based platoon service
  recommendation scheme in {VANET}}.
\newblock \bibinfo{journal}{{IEEE} Trans. Vehicular Technology}
  \bibinfo{volume}{66}, \bibinfo{pages}{1786--1797}.
\bibitem[{Hu et~al.(2011)Hu, Xue, Hong and Wu}]{HuXHW11}
\bibinfo{author}{Hu, W.}, \bibinfo{author}{Xue, K.}, \bibinfo{author}{Hong,
  P.}, \bibinfo{author}{Wu, C.}, \bibinfo{year}{2011}.
\newblock \bibinfo{title}{{ATCS:} {A} novel anonymous and traceable
  communication scheme for vehicular ad hoc networks}.
\newblock \bibinfo{journal}{I. J. Network Security} \bibinfo{volume}{13},
  \bibinfo{pages}{71--78}.
\bibitem[{Iltaf et~al.(2012)Iltaf, Naima, Ghafoor, Abdul, Hussain and
  Mukhtar}]{Iltaf2012Modeling}
\bibinfo{author}{Iltaf}, \bibinfo{author}{Naima}, \bibinfo{author}{Ghafoor},
  \bibinfo{author}{Abdul}, \bibinfo{author}{Hussain},
  \bibinfo{author}{Mukhtar}, \bibinfo{year}{2012}.
\newblock \bibinfo{title}{Modeling interaction using trust and recommendation
  in ubiquitous computing environment}.
\newblock \bibinfo{journal}{Eurasip Journal on Wireless Communications \&
  Networking} \bibinfo{volume}{2012}, \bibinfo{pages}{1--13}.
\bibitem[{Javed et~al.(2018)Javed, Zeadally and Hamid}]{JavedZH18}
\bibinfo{author}{Javed, M.A.}, \bibinfo{author}{Zeadally, S.},
  \bibinfo{author}{Hamid, Z.}, \bibinfo{year}{2018}.
\newblock \bibinfo{title}{Trust-based security adaptation mechanism for
  vehicular sensor networks}.
\newblock \bibinfo{journal}{Computer Networks} \bibinfo{volume}{137},
  \bibinfo{pages}{27--36}.
\bibitem[{Kim and Lee(2014)}]{Kim2014Energy}
\bibinfo{author}{Kim, Y.}, \bibinfo{author}{Lee, S.K.}, \bibinfo{year}{2014}.
\newblock \bibinfo{title}{Energy-efficient wireless hospital sensor networking
  for remote patient monitoring}.
\newblock \bibinfo{journal}{Information Sciences} \bibinfo{volume}{282},
  \bibinfo{pages}{332--349}.
\bibitem[{Lai et~al.(2013)Lai, Lai, Huang and Chao}]{Lai2013Multi}
\bibinfo{author}{Lai, Y.X.}, \bibinfo{author}{Lai, C.F.},
  \bibinfo{author}{Huang, Y.M.}, \bibinfo{author}{Chao, H.C.},
  \bibinfo{year}{2013}.
\newblock \bibinfo{title}{Multi-appliance recognition system with hybrid
  svm/gmm classifier in ubiquitous smart home}.
\newblock \bibinfo{journal}{Information Sciences} \bibinfo{volume}{230},
  \bibinfo{pages}{39--55}.
\bibitem[{Li et~al.(2014)Li, Li, Gao, Zhao, Fan and Han}]{LiLGZFH14}
\bibinfo{author}{Li, Q.}, \bibinfo{author}{Li, Y.}, \bibinfo{author}{Gao, J.},
  \bibinfo{author}{Zhao, B.}, \bibinfo{author}{Fan, W.}, \bibinfo{author}{Han,
  J.}, \bibinfo{year}{2014}.
\newblock \bibinfo{title}{Resolving conflicts in heterogeneous data by truth
  discovery and source reliability estimation}, in: \bibinfo{booktitle}{Int.
  Conf. on Management of Data}, pp. \bibinfo{pages}{1187--1198}.
\bibitem[{Li et~al.(2016)Li, Li, Gao, Su, Zhao, Fan and Han}]{LiLGSZFH16}
\bibinfo{author}{Li, Y.}, \bibinfo{author}{Li, Q.}, \bibinfo{author}{Gao, J.},
  \bibinfo{author}{Su, L.}, \bibinfo{author}{Zhao, B.}, \bibinfo{author}{Fan,
  W.}, \bibinfo{author}{Han, J.}, \bibinfo{year}{2016}.
\newblock \bibinfo{title}{Conflicts to harmony: {A} framework for resolving
  conflicts in heterogeneous data by truth discovery}.
\newblock \bibinfo{journal}{{IEEE} Trans. Knowl. Data Eng.}
  \bibinfo{volume}{28}, \bibinfo{pages}{1986--1999}.
\bibitem[{Liu et~al.(2016)Liu, Peng, Wang, Liu and Huang}]{LiuPWLH16}
\bibinfo{author}{Liu, L.}, \bibinfo{author}{Peng, Y.}, \bibinfo{author}{Wang,
  S.}, \bibinfo{author}{Liu, M.}, \bibinfo{author}{Huang, Z.},
  \bibinfo{year}{2016}.
\newblock \bibinfo{title}{Complex activity recognition using time series
  pattern dictionary learned from ubiquitous sensors}.
\newblock \bibinfo{journal}{Inf. Sci.} \bibinfo{volume}{340-341},
  \bibinfo{pages}{41--57}.
\bibitem[{Lu et~al.(2017)Lu, Heung, Lashkari and Ghorbani}]{LuHLG17}
\bibinfo{author}{Lu, R.}, \bibinfo{author}{Heung, K.},
  \bibinfo{author}{Lashkari, A.H.}, \bibinfo{author}{Ghorbani, A.A.},
  \bibinfo{year}{2017}.
\newblock \bibinfo{title}{A lightweight privacy-preserving data aggregation
  scheme for fog computing-enhanced iot}.
\newblock \bibinfo{journal}{{IEEE} Access} \bibinfo{volume}{5},
  \bibinfo{pages}{3302--3312}.
\bibitem[{Ma et~al.(2006)Ma, Zhao, Chaudhary, Cheng, Yang, Huang and
  Jin}]{MaZCCYHJ06}
\bibinfo{author}{Ma, J.}, \bibinfo{author}{Zhao, Q.},
  \bibinfo{author}{Chaudhary, V.}, \bibinfo{author}{Cheng, J.},
  \bibinfo{author}{Yang, L.T.}, \bibinfo{author}{Huang, R.},
  \bibinfo{author}{Jin, Q.}, \bibinfo{year}{2006}.
\newblock \bibinfo{title}{Ubisafe computing: Vision and challenges {(I)}}, in:
  \bibinfo{booktitle}{Autonomic and Trusted Computing, Third International
  Conference, {ATC} 2006, Wuhan, China, September 3-6, 2006, Proceedings}, pp.
  \bibinfo{pages}{386--397}.
\bibitem[{Miao et~al.(2015)Miao, Jiang, Su, Li, Guo, Qin, Xiao, Gao and
  Ren}]{MiaoJSLGQXGR15}
\bibinfo{author}{Miao, C.}, \bibinfo{author}{Jiang, W.}, \bibinfo{author}{Su,
  L.}, \bibinfo{author}{Li, Y.}, \bibinfo{author}{Guo, S.},
  \bibinfo{author}{Qin, Z.}, \bibinfo{author}{Xiao, H.}, \bibinfo{author}{Gao,
  J.}, \bibinfo{author}{Ren, K.}, \bibinfo{year}{2015}.
\newblock \bibinfo{title}{Cloud-enabled privacy-preserving truth discovery in
  crowd sensing systems}, in: \bibinfo{booktitle}{Proceedings of the 13th {ACM}
  Conference on Embedded Networked Sensor Systems, SenSys 2015, Seoul, South
  Korea, November 1-4, 2015}, pp. \bibinfo{pages}{183--196}.
\bibitem[{Miao et~al.(2017)Miao, Su, Jiang, Li and Tian}]{MiaoSJLT17}
\bibinfo{author}{Miao, C.}, \bibinfo{author}{Su, L.}, \bibinfo{author}{Jiang,
  W.}, \bibinfo{author}{Li, Y.}, \bibinfo{author}{Tian, M.},
  \bibinfo{year}{2017}.
\newblock \bibinfo{title}{A lightweight privacy-preserving truth discovery
  framework for mobile crowd sensing systems}, in: \bibinfo{booktitle}{Proc. of
  INFOCOM}, pp. \bibinfo{pages}{1--9}.
\bibitem[{Nieto-Reyes et~al.(2017)Nieto-Reyes, Duque, Montaã±A and
  Lage}]{Nieto2017Classification}
\bibinfo{author}{Nieto-Reyes, A.}, \bibinfo{author}{Duque, R.},
  \bibinfo{author}{Montaã±A, J.L.}, \bibinfo{author}{Lage, C.},
  \bibinfo{year}{2017}.
\newblock \bibinfo{title}{Classification of alzheimer's patients through
  ubiquitous computing}.
\newblock \bibinfo{journal}{Sensors} \bibinfo{volume}{17},
  \bibinfo{pages}{1679}.
\bibitem[{Paillier(1999)}]{Paillier99}
\bibinfo{author}{Paillier, P.}, \bibinfo{year}{1999}.
\newblock \bibinfo{title}{Public-key cryptosystems based on composite degree
  residuosity classes}, in: \bibinfo{booktitle}{Advances in Cryptology -
  {EUROCRYPT} '99, International Conference on the Theory and Application of
  Cryptographic Techniques, Prague, Czech Republic, May 2-6, 1999, Proceeding},
  pp. \bibinfo{pages}{223--238}.
\bibitem[{Patwardhan et~al.(2006)Patwardhan, Joshi, Finin and
  Yesha}]{PatwardhanJFY06}
\bibinfo{author}{Patwardhan, A.}, \bibinfo{author}{Joshi, A.},
  \bibinfo{author}{Finin, T.}, \bibinfo{author}{Yesha, Y.},
  \bibinfo{year}{2006}.
\newblock \bibinfo{title}{A data intensive reputation management scheme for
  vehicular ad hoc networks}, in: \bibinfo{booktitle}{Proc. of Int. Conf. on
  Mobile and Ubiquitous Systems: Computing, Networking and Services}, pp.
  \bibinfo{pages}{1--8}.
\bibitem[{Raya et~al.(2008)Raya, Papadimitratos, Gligor and Hubaux}]{RayaPGH08}
\bibinfo{author}{Raya, M.}, \bibinfo{author}{Papadimitratos, P.},
  \bibinfo{author}{Gligor, V.D.}, \bibinfo{author}{Hubaux, J.},
  \bibinfo{year}{2008}.
\newblock \bibinfo{title}{On data-centric trust establishment in ephemeral ad
  hoc networks}, in: \bibinfo{booktitle}{Proc. of IEEE INFOCOM}, pp.
  \bibinfo{pages}{1238--1246}.
\bibitem[{Sang et~al.(2009)Sang, Shen and Tian}]{SangST09}
\bibinfo{author}{Sang, Y.}, \bibinfo{author}{Shen, H.}, \bibinfo{author}{Tian,
  H.}, \bibinfo{year}{2009}.
\newblock \bibinfo{title}{Privacy-preserving tuple matching in distributed
  databases}.
\newblock \bibinfo{journal}{{IEEE} Trans. Knowl. Data Eng.}
  \bibinfo{volume}{21}, \bibinfo{pages}{1767--1782}.
\bibitem[{Vendramin et~al.(2012)Vendramin, Munaretto, Delgado and
  Viana}]{Vendramin2012GrAnt}
\bibinfo{author}{Vendramin, A.C.K.}, \bibinfo{author}{Munaretto, A.},
  \bibinfo{author}{Delgado, M.R.}, \bibinfo{author}{Viana, A.C.},
  \bibinfo{year}{2012}.
\newblock \bibinfo{title}{Grant: Inferring best forwarders from complex
  networks’ dynamics through a greedy ant colony optimization}.
\newblock \bibinfo{journal}{Computer Networks} \bibinfo{volume}{56},
  \bibinfo{pages}{997--1015}.
\bibitem[{Xia et~al.(2015)Xia, Liu, Li, Ahmed, Yang and Ma}]{XiaLLAYM15}
\bibinfo{author}{Xia, F.}, \bibinfo{author}{Liu, L.}, \bibinfo{author}{Li, J.},
  \bibinfo{author}{Ahmed, A.M.}, \bibinfo{author}{Yang, L.T.},
  \bibinfo{author}{Ma, J.}, \bibinfo{year}{2015}.
\newblock \bibinfo{title}{{BEEINFO:} interest-based forwarding using artificial
  bee colony for socially aware networking}.
\newblock \bibinfo{journal}{{IEEE} Trans. Vehicular Technology}
  \bibinfo{volume}{64}, \bibinfo{pages}{1188--1200}.
\bibitem[{Xiao et~al.(2007a)Xiao, Du, Zhang, Hu and Guizani}]{0001DZHG07}
\bibinfo{author}{Xiao, Y.}, \bibinfo{author}{Du, X.}, \bibinfo{author}{Zhang,
  J.}, \bibinfo{author}{Hu, F.}, \bibinfo{author}{Guizani, S.},
  \bibinfo{year}{2007}a.
\newblock \bibinfo{title}{Internet protocol television {(IPTV):} the killer
  application for the next-generation internet}.
\newblock \bibinfo{journal}{{IEEE} Communications Magazine}
  \bibinfo{volume}{45}, \bibinfo{pages}{126--134}.
\bibitem[{Xiao et~al.(2007b)Xiao, Rayi, Sun, Du, Hu and Galloway}]{XiaoRSDHG07}
\bibinfo{author}{Xiao, Y.}, \bibinfo{author}{Rayi, V.K.}, \bibinfo{author}{Sun,
  B.}, \bibinfo{author}{Du, X.}, \bibinfo{author}{Hu, F.},
  \bibinfo{author}{Galloway, M.}, \bibinfo{year}{2007}b.
\newblock \bibinfo{title}{A survey of key management schemes in wireless sensor
  networks}.
\newblock \bibinfo{journal}{Computer Communications} \bibinfo{volume}{30},
  \bibinfo{pages}{2314--2341}.
\bibitem[{Xing and Julien(2013)}]{XingJ13}
\bibinfo{author}{Xing, M.}, \bibinfo{author}{Julien, C.}, \bibinfo{year}{2013}.
\newblock \bibinfo{title}{Trust-based, privacy-preserving context aggregation
  and sharing in mobile ubiquitous computing}, in: \bibinfo{booktitle}{Mobile
  and Ubiquitous Systems: Computing, Networking, and Services - 10th
  International Conference, {MOBIQUITOUS} 2013, Tokyo, Japan, December 2-4,
  2013, Revised Selected Papers}, pp. \bibinfo{pages}{316--329}.
\bibitem[{Xu et~al.(2017)Xu, Lu, Wang, Zhu and Huang}]{XuLWZH17}
\bibinfo{author}{Xu, C.}, \bibinfo{author}{Lu, R.}, \bibinfo{author}{Wang, H.},
  \bibinfo{author}{Zhu, L.}, \bibinfo{author}{Huang, C.}, \bibinfo{year}{2017}.
\newblock \bibinfo{title}{{TJET:} ternary join-exit-tree based dynamic key
  management for vehicle platooning}.
\newblock \bibinfo{journal}{{IEEE} Access} \bibinfo{volume}{5},
  \bibinfo{pages}{26973--26989}.
\bibitem[{Xu et~al.(2018)Xu, Xue, Yang and Hong}]{XuXYH18}
\bibinfo{author}{Xu, J.}, \bibinfo{author}{Xue, K.}, \bibinfo{author}{Yang,
  Q.}, \bibinfo{author}{Hong, P.}, \bibinfo{year}{2018}.
\newblock \bibinfo{title}{{PSAP:} pseudonym-based secure authentication
  protocol for {NFC} applications}.
\newblock \bibinfo{journal}{{IEEE} Trans. Consumer Electronics}
  \bibinfo{volume}{64}, \bibinfo{pages}{83--91}.
\bibitem[{Xue et~al.(2018)Xue, Hong, Ma, Wei, Hong and Yu}]{XueHMWHY18}
\bibinfo{author}{Xue, K.}, \bibinfo{author}{Hong, J.}, \bibinfo{author}{Ma,
  Y.}, \bibinfo{author}{Wei, D.S.L.}, \bibinfo{author}{Hong, P.},
  \bibinfo{author}{Yu, N.}, \bibinfo{year}{2018}.
\newblock \bibinfo{title}{Fog-aided verifiable privacy preserving access
  control for latency-sensitive data sharing in vehicular cloud computing}.
\newblock \bibinfo{journal}{{IEEE} Network} \bibinfo{volume}{32},
  \bibinfo{pages}{7--13}.
\bibitem[{Yang et~al.(2019)Yang, Xue, Xu, Wang, Li and Yu}]{YangXXWLY19}
\bibinfo{author}{Yang, Q.}, \bibinfo{author}{Xue, K.}, \bibinfo{author}{Xu,
  J.}, \bibinfo{author}{Wang, J.}, \bibinfo{author}{Li, F.},
  \bibinfo{author}{Yu, N.}, \bibinfo{year}{2019}.
\newblock \bibinfo{title}{Anfra: Anonymous and fast roaming authentication for
  space information network}.
\newblock \bibinfo{journal}{{IEEE} Trans. Information Forensics and Security}
  \bibinfo{volume}{14}, \bibinfo{pages}{486--497}.
\bibitem[{Zhang et~al.(2019)Zhang, Zhu, Xu, Sharif, Du and
  Guizani}]{ZhangZXSDG19}
\bibinfo{author}{Zhang, C.}, \bibinfo{author}{Zhu, L.}, \bibinfo{author}{Xu,
  C.}, \bibinfo{author}{Sharif, K.}, \bibinfo{author}{Du, X.},
  \bibinfo{author}{Guizani, M.}, \bibinfo{year}{2019}.
\newblock \bibinfo{title}{{LPTD:} achieving lightweight and privacy-preserving
  truth discovery in ciot}.
\newblock \bibinfo{journal}{Future Generation Comp. Syst.}
  \bibinfo{volume}{90}, \bibinfo{pages}{175--184}.
\bibitem[{Zhang et~al.(2018)Zhang, Wang and Wang}]{ZhangWW18}
\bibinfo{author}{Zhang, M.}, \bibinfo{author}{Wang, J.}, \bibinfo{author}{Wang,
  W.}, \bibinfo{year}{2018}.
\newblock \bibinfo{title}{Heterank: {A} general similarity measure in
  heterogeneous information networks by integrating multi-type relationships}.
\newblock \bibinfo{journal}{Inf. Sci.} \bibinfo{volume}{453},
  \bibinfo{pages}{389--407}.
\bibitem[{Zhitao~Guan and Yu(2019)}]{Guan}
\bibinfo{author}{Zhitao~Guan, Yue~Zhang, L.Z.L.W.}, \bibinfo{author}{Yu, S.},
  \bibinfo{year}{2019}.
\newblock \bibinfo{title}{Effect: an efficient flexible privacy-preserving data
  aggregation scheme with authentication in smart grid}.
\newblock \bibinfo{journal}{Science China Information Sciences}
  \bibinfo{volume}{62}, \bibinfo{pages}{1--14}.
\bibitem[{Zhu et~al.(a)Zhu, Meng, Zhang and Zhan}]{ZhuASAP}
\bibinfo{author}{Zhu, L.}, \bibinfo{author}{Meng, L.}, \bibinfo{author}{Zhang,
  Z.}, \bibinfo{author}{Zhan, Q.}, a.
\newblock \bibinfo{title}{Asap: An anonymous smart-parking and payment scheme
  in vehicular networks}.
\newblock \bibinfo{journal}{IEEE Transactions on Dependable \& Secure
  Computing} \bibinfo{volume}{PP}, \bibinfo{pages}{1--1}.
\bibitem[{Zhu et~al.(2019)Zhu, Wu, Gai and Choo}]{ZhuWGC19}
\bibinfo{author}{Zhu, L.}, \bibinfo{author}{Wu, Y.}, \bibinfo{author}{Gai, K.},
  \bibinfo{author}{Choo, K.R.}, \bibinfo{year}{2019}.
\newblock \bibinfo{title}{Controllable and trustworthy blockchain-based cloud
  data management}.
\newblock \bibinfo{journal}{Future Generation Comp. Syst.}
  \bibinfo{volume}{91}, \bibinfo{pages}{527--535}.
\bibitem[{Zhu et~al.(b)Zhu, Zhang, Xu, Du, Xu, Sharif and Guizani}]{zhuzxdxsm}
\bibinfo{author}{Zhu, L.}, \bibinfo{author}{Zhang, C.}, \bibinfo{author}{Xu,
  C.}, \bibinfo{author}{Du, X.}, \bibinfo{author}{Xu, R.},
  \bibinfo{author}{Sharif, K.}, \bibinfo{author}{Guizani, M.}, b.
\newblock \bibinfo{title}{{PRIF:} {A} privacy-preserving interest-based
  forwarding scheme for social internet of vehicles}.
\newblock \bibinfo{journal}{IEEE Internet of Things Journal}
  \bibinfo{volume}{5}, \bibinfo{pages}{2457--2466}.
\bibitem[{Zhu et~al.(2018)Zhu, Zhang, Xu and Sharif}]{ZhuZXS18}
\bibinfo{author}{Zhu, L.}, \bibinfo{author}{Zhang, C.}, \bibinfo{author}{Xu,
  C.}, \bibinfo{author}{Sharif, K.}, \bibinfo{year}{2018}.
\newblock \bibinfo{title}{Rtsense: Providing reliable trust-based crowdsensing
  services in {CVCC}}.
\newblock \bibinfo{journal}{{IEEE} Network} \bibinfo{volume}{32},
  \bibinfo{pages}{20--26}.

\end{thebibliography}

\end{document}